\documentclass[NewJournal,InsideFigs,SingleSpace,11pt,NoLists,NoLineNumbers]{ascelike_rev}

\usepackage[T1]{fontenc}
\usepackage[latin9]{inputenc}
\usepackage{amsmath}
\usepackage{amssymb}
\usepackage{graphicx}
\usepackage{esint}
\usepackage{enumitem}
\usepackage{subfig}
\begin{document}
\title{\Large Contact transience during slow loading\\
        of dense granular materials}
\author{
Matthew R. Kuhn%
%
\thanks{%
Professor, Dept.\ of Civil Engrg.,
Donald P.\ Shiley School of Engrg., Univ.\ of Portland,
5000 N.\ Willamette Blvd.,
Portland, OR  97203 (corresponding author). E-mail: kuhn@up.edu.},
\ M.~ASCE}
%
%
\maketitle
\begin{abstract}
The irregularity of particle motions during quasi-static deformation
is investigated using discrete element (DEM) simulations of
of sphere and sphere-cluster assemblies.
Three types of inter-particle movements are analyzed:
relative motions of particle centers,
relative motions of material points of two particles at their contact,
and the traversal of contacts across the surfaces of particles.
Motions are a complex combination of rolling, sliding, and elastic
distortion at the contacts, and
all motions are highly irregular and variant, qualities that
increase with increasing strain.
The relative motions of particle centers diverges
greatly from those of an affine displacement of the particles.
The motions of the non-convex sphere-cluster particles was more regular
that those of spheres.
The paper also investigates the effect of the distance between
two remote particles and their pair-wise relative displacements.
Even for particle-pairs separated by more than 6 intermediate particles,
the relative motions do not conform with the mean deformation (affine)
field.
Force chains are shown to be transient features, which survive
only briefly across elapsed strains.
\end{abstract}
\KeyWords{Granular material, micro-mechanics, discrete element method,
          contact mechanics, simulation.}
\section{Introduction}
Two approaches predominate the
micro-mechanical modeling of granular materials.
In one approach, the grains are viewed as moving and interacting 
in an orderly, regular manner, 
respecting simple kinematic or statical fields.
As an example, introductory geotechnical textbooks often represent
soil failure as a continuous planar sheet of particles sliding across a
facing sheet of particles, and dilation is often
represented as the opening
of pore space between the two sheets.
Although most researchers now discount such simple views as being
rudimentary,
current multi-scale analyses of granular material
often assume a simple
spatial homogenization of the underlying micro-scale complexity
and also assume that the temporal transitioning of the 
micro-scale landscape
occurs smoothly and continuously
through a continuum of states as the material
undergoes bulk deformation.
The loading of granular materials is now known to involve the
interactions of particles that individually progress in a
seemingly disordered and random fashion, but in a manner
that collectively produces
repeatable and consistent progressions of deformation and stress.
The paper considers this second view of granular materials and quantifies
the non-uniform and seemingly capricious, erratic nature of grain
interactions.
Whereas
previous studies have primarily examined
the spatial domain and have focused upon spatial irregularities
and heterogeneities (shear bands, micro-bands, topological irregularities,
etc.), the paper examines the temporal domain and seeks to
characterize the transience of grain movements during
deviatoric loading.
\par
The paper considers the rate at which the particles interact
and are rearranged during quasi-static loading,
by addressing the following questions:
\begin{itemize}
\item
Is contact motion regular,
and how severe is any irregularity in this motion?
For example,
while a pair of grains is in contact, does their point of
contact shift across the surfaces of the two grains in a smooth,
regular manner?
\item
To what extent do the contact motions conform with the imposed
bulk, mean deformation field?
At what length scale do particle movements begin
to approach compliance with the mean field?
\item
Loading behavior is known to originate in the interactions of
touching grains.
How long do grains typically remain in contact?
At what rate are contacts created and expired during
quasi-static loading?
\item
In a recent view of granular stress, the contacts are partitioned
into two sets: those bearing a force greater than the mean
(the ``strong'' contacts) and those that support a smaller force
(``weak'' contacts).
What is the longevity of this partition,
and how frequently do contacts pass
between the two sets?
\item
Load-bearing chains of particles, ``force chains,'' are the most
apparent feature of force transmission within granular materials.
Over what time-scale are these force chains persistent?
\end{itemize}
Many of these questions have already received attention, as will be
cited below.
The questions are explored with 3D discrete element simulations
using assemblies of spherical particles and assemblies of non-convex 
particles --- agglomerated clusters of spheres termed ``sphere-clusters.''
The quasi-static
simulations reveal the non-smooth, transient, and inconstant
nature of the grain motions and inter-granular forces,
which occurs to a degree that is quite remarkable.
We begin with a brief description of the simulation methods
and recount general trends in the particle movements.
We then answer, in turn, the questions posed above, quantifying
the transient and irregular character of the micro-scale granular landscape.
We conclude with perspectives on the results and
suggestions of possible future approaches to
micro-mechanical analysis.
\section{DEM Methods}
\begin{figure}
  \centering
  \includegraphics{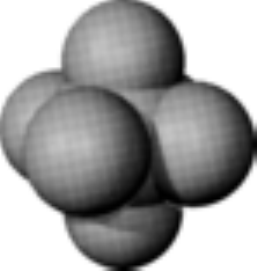}
  \caption{Non-convex particle: a bonded sphere-cluster.
           \label{fig:bumpy}}
\end{figure}

The discrete element method (DEM) was used to simulate the loading
of two assemblies: one composed of sphere particles
and the other of sphere-clusters.
The clusters were formed from seven overlapping spheres, having
a larger central particle and six smaller satellite particles
in an octahedral arrangement
(Fig.~\ref{fig:bumpy}).
Cluster particles were used to achieve a more realistic
representation of natural materials
by incorporating a modest non-convexity
and to compare the micro-scale behaviors of convex (sphere)
and non-convex particles.
These sphere-clusters can be efficiently implemented
and have been shown to produce results that are very similar
to those of a poorly graded natural sands
composed of fine sub-rounded
particles \cite{Kuhn:2014c}.
The range of particle sizes was small,
between 0.075~mm and 0.28~mm,
with a median size $D_{50}$ of 0.165~mm.
The two assemblies of 6400 particles were isotropically
compacted within periodic boundaries
to a medium density (void ratios of 0.576 and 0.706 for the spheres
and sphere-clusters).
The assemblies are large enough to
capture the average material behavior but sufficiently small to
prevent meso-scale localization, such as shear bands.
\par
The contact model was a full implementation of a Hertz-Mindlin
contact between elastic-frictional spheres by using the
J\"{a}ger algorithm, which can model arbitrary sequences
of normal and tangential contact movements \cite{Kuhn:2011a}.
The simulations were conducted with
an inter-particle friction coefficient 
$\mu=0.60$,
particle shear modulus $G=29$~GPa, and Poisson ratio $\nu=0.15$.
\par
\begin{figure}
  \centering
  \mbox{%
    \subfloat[Triaxial compression]{\includegraphics{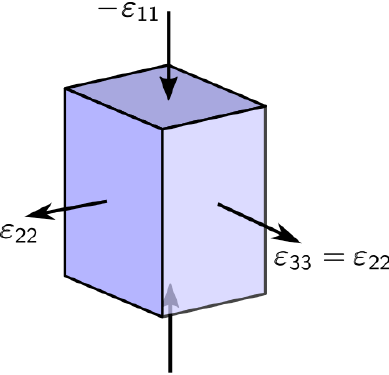}}\quad
    \subfloat[Unit sphere]{\includegraphics{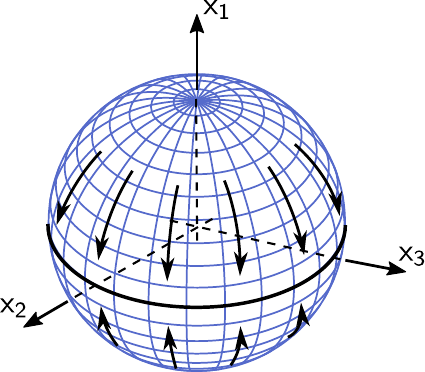}}\quad
    \subfloat[Unit sphere]{\includegraphics{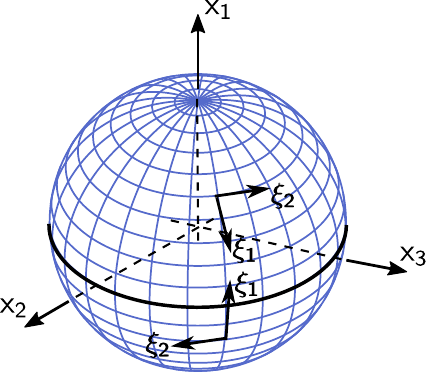}}
  }
  \caption{Assembly loading and movements of contact normals:
           (a) triaxial compression loading of the current DEM simulations,
           (b) unit sphere showing general conveyance of
               contact normals during triaxial
               compression, and 
           (c) local coordinate system 
               $\xi_{1}$--$\xi_{2}$ for the tangential movements
               of contacts.
           \label{fig:triax}}
\end{figure}

The two assemblies started with an initial isotropic arrangement of
particles, confined with an isotropic stress
of 100~kPa.
The assemblies were then monotonically loaded under triaxial conditions
in which the assembly height (in direction $x_{1}$) was reduced
at a constant slow rate ($\dot{\varepsilon}_{11}<0$)
while maintaining a constant mean stress
$p$ by adjusting the transverse, horizontal widths
in directions $x_{2}$ and $x_{3}$,
with $\dot{\varepsilon}_{22}=\dot{\varepsilon}_{33}$.
These conditions are represented in Fig.~\ref{fig:triax}a.
\citeN{daCruz:2005a} classify granular flows in three
regimes, depending upon the rapidity of deformation:
quasi-static (slow), collisional, and an intermediate, dense regime of
flow.
The current simulations are quasi-static, with an inertial number
$I=\dot{\gamma}\sqrt{m/(pd)}$ of less than 0.0003.
The internal kinetic energy of the assembly during loading was
consistently less than 3\% of the internal elastic energy throughout
the simulation, and the internal kinetic energy was less than
0.3\% of the work expended in frictional sliding for each 1\% of strain.
The particles remained in near-equilibrium throughout the simulations,
with an average imbalance of force that was less than 2\% of the average
contact force.
\par
\begin{figure}
\centering
\mbox{%
  \subfloat[]{\includegraphics{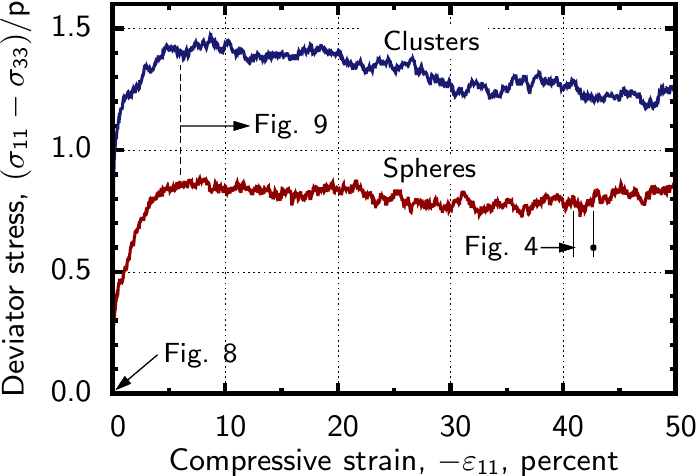}}\qquad
  \subfloat[]{\includegraphics{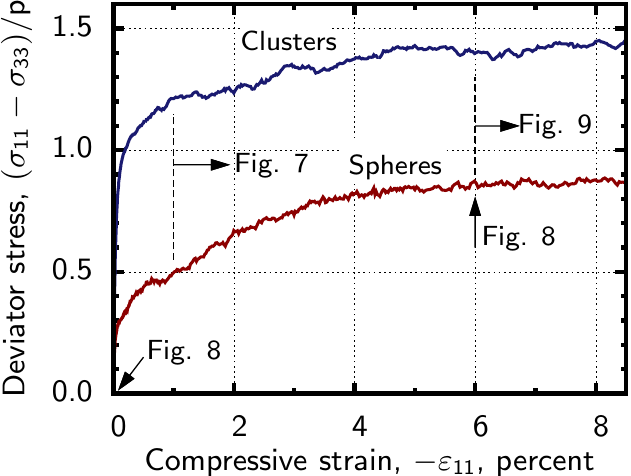}}
}
\caption{Stress-strain behavior of two assemblies in triaxial compression.
         \label{fig:stress}}
\end{figure}

Figure~\ref{fig:stress} shows the stress-strain behavior of the assemblies.
As expected, the assemblies of non-convex sphere-clusters are
considerably stronger and stiffer than the sphere assemblies
(peak angles of internal friction $\phi$
of 22.7$^{\circ}$ and 36.4$^{\circ}$ respectively).
\section{Grain Motions}
In recent work by the author on the loading of granular media,
the \emph{averaged} contact motions were found to follow a regular trend during
bulk deformation:
contacts tend to migrate from orientations in which the principal
strain is compressive and toward orientations
of extensional principal strains \cite{Kuhn:2010a}.
That is, if one considers all
contact orientations as populating the unit sphere,
the contacts are, on average, conveyed from directions of compressive
strain toward directions of extension.
For the triaxial compression constant-mean-stress
conditions of the current simulations (Fig.~\ref{fig:triax}a),
the general motion of contacting particle pairs is a tangential
shifting of the contacts' normal vectors from the compression
direction (in Fig.~\ref{fig:triax}b, the north-south $x_{1}$
and $-x_{1}$ directions) toward the transverse directions of expansion
(toward the $x_{2}$--$x_{3}$ equator in Fig.~\ref{fig:triax}b).
As will be seen, the individual contact motions are quite
diverse --- nearly erratic --- and
short-lived, to the extent that this model of general
contact conveyance is only revealed
by averaging the motions of thousands of contacts.
The paper focuses, instead, on the diverse, fleeting, and transitory
aspect of granular motion.
\par
We will characterize the nature and diversity of contact
movements in three respects,
\begin{enumerate}[label=(\alph*\upshape)]
\item
as relative movements of the centers
of particles that form contacting pairs,
since these movements are the micro-mechanical source of
bulk strain \cite{Bagi:1996a,Satake:2004a},
\item
as the relative movements of
particles' material points in the vicinities of their contacts,
since these movements
are the source of the evolving contact forces and
of the bulk stress and stiffness
\cite{Christoffersen:1981a,Satake:2001a}, and
\item
as movements of the contact points (actually of the
contact zones) migrating across the surfaces of particles.
Advanced theories of friction acknowledge the
effects of sliding distance, sliding rate, and heat generation
and the interaction of rolling and sliding,
matters that arise from migrations across contacting surfaces
\cite{DiToro:2004a,Rice:2006a,Kuhn:2014b}.
\end{enumerate}
These three movements are designated as
$d\mathbf{u}^{pq,\text{centers}}$,
$d\mathbf{u}^{pq,\text{rigid}}$,
and
$\frac{1}{2}
(d\mathbf{u}^{pq,\text{roll}} +
d\mathbf{u}^{pq,\text{slip}})$
in the following derivations.

\par
Consider two particles, $p$ and $q$,
that touch at contact $pq$.
If the particles are rigid, the relative incremental
movement of points on either side of the
contact (of material points in $q$ relative to those in $p$)
is
\begin{equation}
d\mathbf{u}^{pq,\text{rigid}}=d\mathbf{u}^{pq,\text{centers}}
+
\left(
d\boldsymbol{\theta}^{q}\times\mathbf{r}^{q} -
d\boldsymbol{\theta}^{p}\times\mathbf{r}^{p}
\right)
\label{eq:durigid}
\end{equation}
where $d\boldsymbol{\theta}^{p}$ and $d\boldsymbol{\theta}^{q}$
are the rotation increments of the two particles;
$\mathbf{r}^{p}$ and $\mathbf{r}^{q}$ are vectors that join
the centers of each particle to the contact point;
and
$d\mathbf{u}^{pq,\text{centers}}$ is the increment
of relative movement of the two particles' centers:
\begin{equation}
d\mathbf{u}^{pq,\text{centers}}=d\mathbf{u}^{q}-d\mathbf{u}^{p}
\label{eq:centers}
\end{equation}
with $d\mathbf{u}^{p}$ and $d\mathbf{u}^{q}$ being
the separate movements of each particle's
center.
\par
If the particles were truly rigid, an arbitrary increment
$d\mathbf{u}^{pq,\text{rigid}}$ could produce inter-penetration
at the contact.
In contrast,
an elastic Hertz contact between \emph{deformable} spherical bodies
produces deformation in the contact vicinity, so that contact occurs
within a circular patch on each surface, with
the radius of this patch dependant on the normal force.
According to Cattaneo-Mindlin theory,
a modest tangential contact motion $d\mathbf{u}^{pq,\text{rigid}}$ produces
micro-slip inside of an annular
peripheral zone within the contact patch, even as the 
particles frictionally adhere within the central
portion of the patch.
With greater movement,
the tangential contact force reaches the frictional limit,
and the center of the contact patch can
then move (slip) across the surfaces of the
two particles.
In this regard,
the rigid increment $d\mathbf{u}^{pq,\text{rigid}}$ can be separated into
two parts: one corresponding to rigid slip (and movement of the contact patch)
and the other an elasto-plastic
(or elasto-frictional) increment that produces deformation
within the particle bodies
and micro-slip
within the contact patch,
\begin{equation}
d\mathbf{u}^{pq,\text{rigid}} = d\mathbf{u}^{pq,\text{slip}} +
                                d\mathbf{u}^{pq,\text{ep}}
\label{eq:rigidparts}
\end{equation}
noting that slip movement
is possible only when the frictional limit
is reached:
when the magnitude of tangential force
$|\mathbf{f}^{pq,\text{t}}|$ reaches the
product of the normal force $f^{pq,\text{n}}$ and the friction
coefficient $\mu$.
\par
We can also separate each increment in Eq.~(\ref{eq:rigidparts}) 
into motions that
are normal and tangential to the contact plane,
denoted with ``n'' and ``t'' superscripts:
\begin{equation}
du^{pq,\circ,\text{n}} =
d\mathbf{u}^{pq,\circ}\cdot \mathbf{n}^{pq}, \quad
d\mathbf{u}^{pq,\circ,\text{t}} =
d\mathbf{u}^{pq,\circ} - du^{pq,\circ,\text{n}} \mathbf{n}^{pq}
\label{eq:normaltangent}
\end{equation}
where $\mathbf{n}^{pq}$ is the unit normal, directed outward
from $p$.
Note that the slip motion
$\mathbf{u}^{pq,\text{slip}}$ in Eq.~(\ref{eq:rigidparts}) is
entirely tangential.
\par
Besides movement $d\mathbf{u}^{pq,\text{rigid}}$,
other types of motion can also cause the contact patch to migrate
across the particles.
Twelve components of motion are available to the two particles
(three components of rotation and displacement for each particle),
and only three components are expressed in 
the objective motion $d\mathbf{u}^{pq,\text{rigid}}$.
Six additional 
components are associated with non-objective, rigid-body motions
\cite{Kuhn:2005c}.
Of the three remaining objective motions, two are associated
with the contact rolling that occurs within the contact's
tangential plane, 
and these two rolling components can cause the contact patch to shift
across the particles' surfaces.
\par
The rolling of two \emph{rigid} particles depends upon their movements and
upon the local shapes of their surfaces at the contact
\cite{Kuhn:2004b}:
\begin{equation}
d\mathbf{u}^{pq,\text{roll}} =
- \left( \mathbf{K}^{p} + \mathbf{K}^{q}\right)^{-1}\cdot
\left[
(d\boldsymbol{\theta}^{q} - d\boldsymbol{\theta}^{p})
\times \mathbf{n}^{pq}
+ \frac{1}{2}
( \mathbf{K}^{p} - \mathbf{K}^{q})\cdot
d\mathbf{u}^{pq,\text{rigid}}
\right]
\label{eq:duroll}
\end{equation}
where $\mathbf{K}^{p}$ and $\mathbf{K}^{q}$ are the surfaces'
curvature tensors.
As two rigid particles roll and slide, they trace paths across
each of their surfaces.
The rolling increment $d\mathbf{u}^{pq,\text{roll}}$ is simply
the \emph{average} increment of two traced paths:
the average of increment
$\mathbf{t}^{p}ds^{p}$ across the surface of $p$, and of increment
$\mathbf{t}^{q}ds^{q}$ across $q$, where the
$\mathbf{t}^{\circ}$ are the
unit directions of these traced paths within the contact's
tangent plane and the $ds^{\circ}$ are the increments' magnitudes.
In a complementary manner, the rigid increment
$d\mathbf{u}^{pq,\text{rigid}}$ of Eq.~(\ref{eq:durigid})
is simply the difference
$\mathbf{t}^{p}ds^{p} - \mathbf{t}^{q}ds^{q}$,
such that the movement of the contact point across the
rigid particle $p$ is
\begin{equation}
\mathbf{t}^{p}ds^{p} =
d\mathbf{u}^{pq,\text{roll}} +
{\textstyle \frac{1}{2}}
d\mathbf{u}^{pq,\text{rigid}}
\label{eq:tds}
\end{equation}
Applying Eq.~(\ref{eq:rigidparts})
for \emph{deformable} particles, the corresponding movement is
\begin{equation}
\mathbf{t}^{p}ds^{p} =
(d\mathbf{u}^{pq,\text{roll}} +
{\textstyle \frac{1}{2}}
d\mathbf{u}^{pq,\text{slip}}) +
\frac{1}{2}
d\mathbf{u}^{pq,\text{ep}}
\label{eq:roll_slip_ep}
\end{equation}
The movement in parentheses is the movement of the center of
the contact patch across the surface of $p$; whereas,
movement $\frac{1}{2}d\mathbf{u}^{pq,\text{ep}}$ is produced
by deformation (within particle $p$ and in the contact's vicinity)
due to frictional adhering.
Although rarely incorporated in DEM simulations
(including the current simulations), the contact force is influenced
by both $d\mathbf{u}^{pq,\text{rigid}}$ and $d\mathbf{u}^{pq,\text{roll}}$,
as an amalgam of sliding friction and rolling friction \cite{Kuhn:2014b}.
\par
We will later compare
the movements at individual contacts
with those that would occur if the particles
conformed perfectly
with the bulk deformation field.
This hypothetical, ``ideal'' 
(affine or mean-field) movement is designated as
$d\mathbf{u}^{pq,\text{ideal}}$ and can also be expressed in its
normal and tangential components:
\begin{gather}
d\mathbf{u}^{pq,\text{ideal}} = 
\mathbf{L}\cdot
(\mathbf{r}^{p}-\mathbf{r}^{q})\,dt \\
du^{pq,\text{ideal,}\,\text{n}}
=
\mathbf{du}^{pq,\text{ideal}}\cdot\mathbf{n}^{pq},\quad
d\mathbf{u}^{pq,\text{ideal,}\,\text{t}} =
d\mathbf{u}^{pq,\text{ideal}}-
du^{pq,\text{ideal,}\,\text{n}}\mathbf{n}^{pq}
\label{eq:pureComponents}
\end{gather}
where $\mathbf{L}dt$ is the bulk Eulerian
strain increment (possibly at large strains)
computed from the displacement gradient: 
$\mathbf{L}dt=\dot{\mathbf{F}}\cdot\mathbf{F}^{-1}dt\approx
d\boldsymbol{\varepsilon}$.
The tangential component of this ideal movement
is aligned with the general direction
of contact conveyance during the bulk deformation,
as was described at the start of this section.
\par
\begin{figure}
  \centering
  \includegraphics{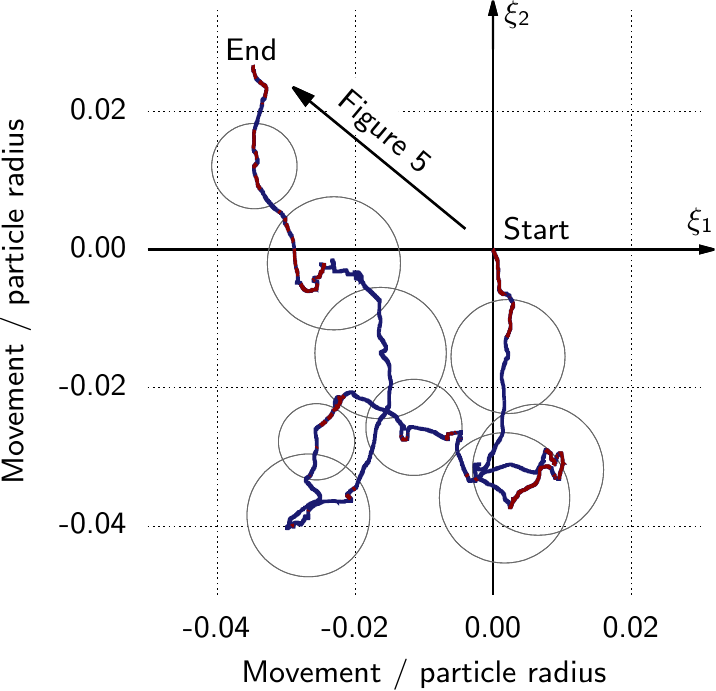}
  \caption{Typical movement of a single contact,
           $d\mathbf{u}^{pq,\text{roll}} +
            {\textstyle \frac{1}{2}}
            d\mathbf{u}^{pq,\text{slip}}$,
            across the surface of one of its particles.
            Circles represent the contact patch.
            \label{fig:ContactMigration}}
\end{figure}

As an example,
Fig.~\ref{fig:ContactMigration}
shows the movement of a single contact
between two spheres recorded during DEM
simulations of triaxial compression
(described in the DEM section and
with Fig.~\ref{fig:triax}).
This contact was
one of about 8500 contacts within the assembly.
The two particles touched when the bulk strain $-\varepsilon_{11}$
was 40.9\% and separated at a bulk strain of 42.7\%.
As will be seen in the next section,
the longevity of this particular contact ($\Delta\varepsilon_{11}=1.8\%$)
was greater than most contacts; but in other respects, this contact was
similar to most of the contacts in the sphere simulation.
Figure~\ref{fig:ContactMigration} shows the tangential
movement of the contact patch,
the motion $d\mathbf{u}^{pq,\text{roll}}+\frac{1}{2}d\mathbf{u}^{pq,\text{slip}}$,
across the surface of one of its particles
(say $p$), projected onto
a $\xi_{1}$--$\xi_{2}$ plane (Fig.~\ref{fig:triax}c).
The $\xi_{1}$ direction corresponds to the direction of general contact conveyance
(i.e., the direction of $d\mathbf{u}^{pq,\text{ideal,}\,\text{t}}$
in Eq.~\ref{eq:pureComponents}$_{2}$).
For triaxial compression in the $x_{1}$ direction,
the $\xi_{1}$ direction is aligned along $x_{1}$ meridians
(the north--south meridians in Fig.~\ref{fig:triax}c), such that
contacts with an orientation in the ``northern hemisphere'' of 
the unit sphere
(i.e. $n_{1}^{pq}>0$) have $\xi_{1}$ directed southward
toward the equator.
For contacts in the southern hemisphere
(i.e. $n_{1}^{pq}<0$), $\xi_{1}$ is directed northward.
\par
As is typical of most contacts,
the contact in Fig.~\ref{fig:ContactMigration}
is seen to wander in a seemingly
erratic manner across the surface of the particle, with much of the motion
in a direction \emph{contrary} to that corresponding to bulk deformation
(i.e., contrary to the mean-field direction of positive $\xi_{1}$).
The contact movement has been normalized by dividing
by the radius of the particle,
so that the movement of this one contact
can be compared with the bulk strain.
If fully unfolded,
the normalized length of this contact's path is
0.256, more than 14 times greater than the elapsed strain of only 0.018.
That is,
when viewed at the micro-scale of contact interactions,
the migration activity of this contact
is much more vigorous than the bulk deformation.
The circles in the figure show sizes of the contact patch,
which is seen to expand and contract as the contact migrates,
corresponding to a growing and diminishing normal force.
Migration of the patch occurs during periods of slip (shown in red)
and also while
the contact is not undergoing slip (shown in blue).
Frictional slip is seen repeatedly to start and stop, 
occurring in a sporadic, fitful sequence of slip episodes.
No aspect of this contact's behavior is regular: force and movement progress
in an irregular, desultory fashion with numerous periods both
of slow, lingering
movement and of darting, rapid movement.
\par
\begin{figure}
  \centering
  \includegraphics{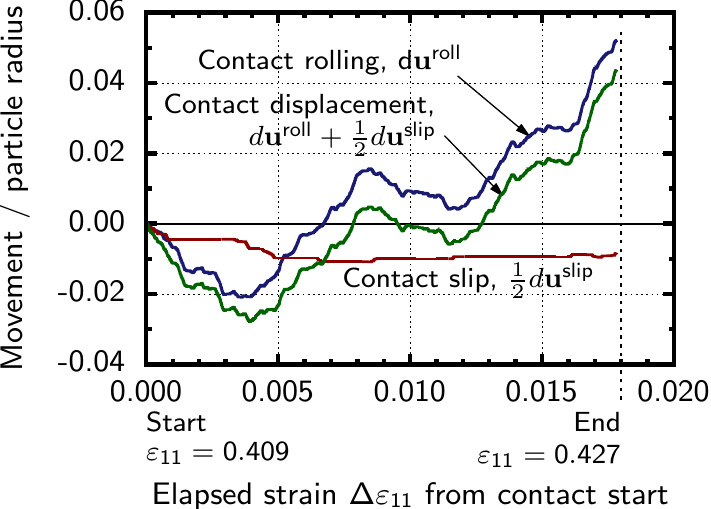}
  \caption{Contact motions of a single particle.
           Results are for the contact shown
           in Fig.\ref{fig:ContactMigration}, projected onto the
           general direction of its movement
           (arrow in Fig.\ref{fig:ContactMigration}).
           \label{fig:onecontact}}
\end{figure}

The general trend of this contact's motion is
indicated by the arrow in Fig.~\ref{fig:ContactMigration}
(this contact lies in the southern hemisphere, and its
general trend is in the southwest direction, contrary
to the northern $\xi_{1}$ direction of general conveyance, see
Fig.~\ref{fig:triax}b).
Figure~\ref{fig:onecontact}
shows a projection of the contact motions in the general direction
of the arrow in Fig.~\ref{fig:ContactMigration}.
The migration of this contact patch is a sum of its
rolling and slip motions
(Eq.~\ref{eq:roll_slip_ep}), which are shown separately in the figure.
Although rolling occurs in a wandering manner
throughout the life-time of this contact,
slip is seen to occur intermittently,
sometimes slipping ``forward'' and sometimes slipping in the
``reverse'' direction.
Oddly, slip and rolling often occur in opposite directions:
at times,
the contact is seen to be rolling forward while slipping backward.
\par
\begin{figure}
  \centering
  \mbox{%
        \subfloat[Strain $-\varepsilon_{11}=0.0025\%$]%
                 {\includegraphics{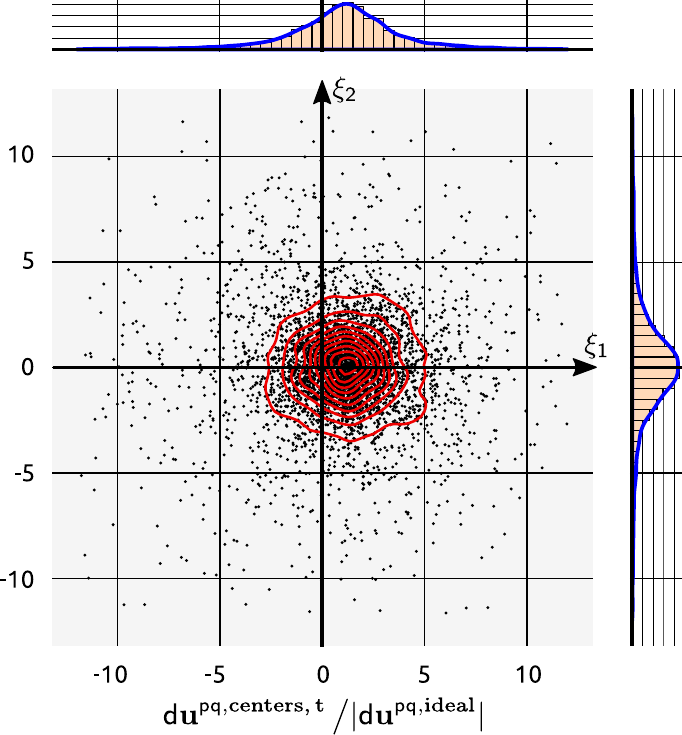}} \quad
        \subfloat[Distributions of $\xi_{1}$ movements at several strains]
                 {\includegraphics{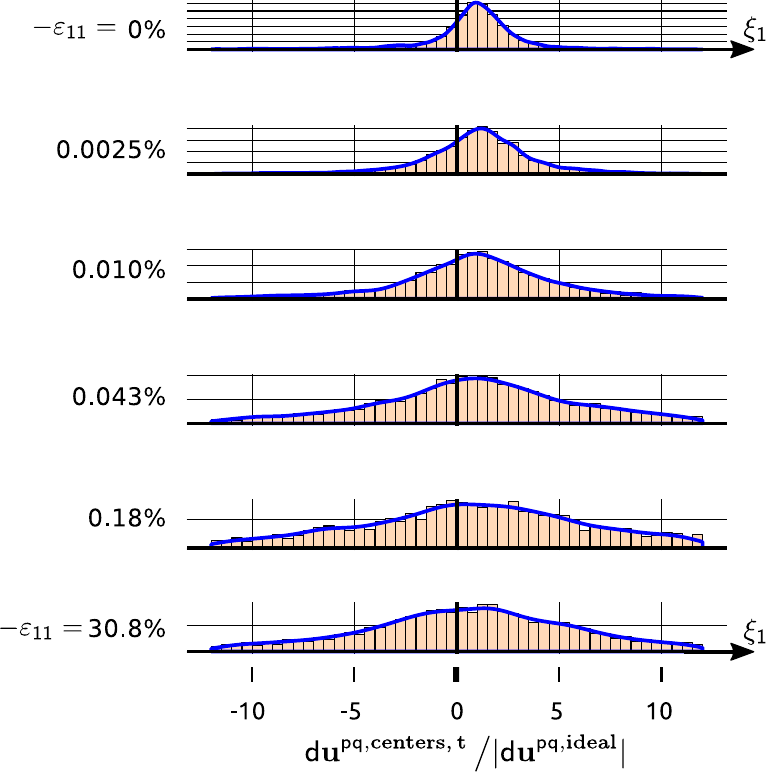}}
  }
  \caption{Contact motions for the sphere assembly:
           (a)~projections of individual motions 
           $d\mathbf{u}^{pq,\text{centers, t}}$ relative to
           the mean-field movement $d\mathbf{u}^{pq,\text{ideal, t}}$,
           (b)~density distributions of movements in
           the $\xi_{1}$ direction of general conveyance
           (see Fig.~\ref{fig:triax}c).
           \label{fig:scatter}}
\end{figure}

The results for this one contact suggest considerable
diversity in particle movements, and we now
consider the motions of a large set of contacts within the assembly.
Figure~\ref{fig:scatter} illustrates the range and nature of motions in the
assembly of spheres shortly after the start of loading, at
strain $-\varepsilon_{11}=0.0025\%$.
The figure shows the results for the
subset of 4630 contacts that lie within the latitudes
of 20$^{\circ}$ and 70$^{\circ}$,
for reasons that are given below.
The scatter plot presents the relative
tangential motions of the centers
of contacting particles, $d\mathbf{u}^{pq,\text{centers,}\,\text{t}}$,
as in Eq.~(\ref{eq:centers}).
In this figure, the motions are projected onto the $\xi_{1}$--$\xi_{2}$
tangential plane, with the positive direction $\xi_{1}$ corresponding
to the expected direction of contact conveyance
(Fig.~\ref{fig:triax}c).
The movements have also been normalized by dividing by the magnitudes
of the tangential
movements that would occur if the pairs of particles had complied
perfectly with the bulk deformation field 
(the distance $|d\mathbf{u}^{pq,\text{ideal,}\,\text{t}}|$, as
in Eq.~\ref{eq:pureComponents}$_{2}$); that is, the
movement at each contact has been divided by the corresponding quantity
$|d\mathbf{u}^{pq,\text{ideal,}\,\text{t}}|$,
which is the magnitude of movement if the two particles
conformed to the mean, uniform deformation field
(i.e. affine motion).
If every particle complied with the mean field, all values would
lie at the single point $(1,0)$ in the $\xi_{1}$--$\xi_{2}$ plane.
Because the magnitude of the ideal motion approaches zero at the
poles and at the equator (i.e., when any of the
$n_{i}^{pq}$ components is close to zero, Fig.~\ref{fig:triax}b),
we restrict Fig.~\ref{fig:scatter}
to the subset of particles that lie in the
``temperate zone'', between latitudes 20$^{\circ}$
and 70$^{\circ}$ (again, Fig.~\ref{fig:triax}b).
\par
In regard to the motions of the particles' centers (Fig.~\ref{fig:scatter}),
the movements are quite varied and can diverge greatly from the mean field.
Even at the very small strain presented in the scatter plot
($-\varepsilon_{11}=0.0025\%$),
some particles move at rates many times faster than those
of the mean field and in directions that depart from the
ideal direction (positive $\xi_{1}$ in the figure),
with many moving in the the opposite, ``wrong''  direction.
The density plots on the right of Fig.~\ref{fig:scatter}
show the distribution of contact motions in the $\xi_{1}$ direction,
and we see that
the diversity of motion increases with increasing strain.
At both small and large strains, we can discern a slight bias
in the contact motions toward the direction of positive $\xi_{1}$,
which is
the direction of general particle conveyance,
but this bias (of about 1) is small when compared with the dispersion of
the motions and can only be distinguished by examining several thousands
of contacts. 
%
\par
The variable nature of the tangential contact motions is also displayed in
Table~\ref{table:std1} for the assemblies of spheres and of non-convex
sphere-clusters, at both small and large strains and for various
types of motion.
The table gives the standard deviations of various particle motions,
after they have been normalized by dividing by the reference tangential
motion $|d\mathbf{u}^{pq,\text{ideal,}\,\text{t}}|$.
Only contacts with orientations between latitudes
20$^{\circ}$ and 70$^{\circ}$ are reported in the table.
\begin{table}
\centering\small
\caption{\small
         Standard deviations of tangential
         contact movements. All movements
         are divided by the mean-field tangential movement
         $|d\mathbf{u}^{pq,\text{ideal,}\,\text{t}}|$
         of Eq.~\ref{eq:pureComponents}$_{2}$.
         \label{table:std1}}
\begin{tabular}{lccc}
\hline
 & & \multicolumn{2}{c}{Particle shape}\\ \cline{3-4}
 & &         & Sphere-\\
 & & Spheres & clusters\\
 \hline
Small strains ($-\varepsilon_{11}<0.01\%$) & & & \\
\quad$\text{std}(d\mathbf{u}^{pq,\text{centers, t}}/|d\mathbf{u}^{pq,\text{ideal,}\,\text{t}}|)$ & Eq.~(\ref{eq:durigid}) &  7.4 & 5.4 \\
\quad$\text{std}(d\mathbf{u}^{pq,\text{rigid, t}}/|d\mathbf{u}^{pq,\text{ideal,}\,\text{t}}|)$ & Eq.~(\ref{eq:centers})   &  5.9 & 5.4 \\
\quad$\text{std}(d\mathbf{u}^{pq,\text{roll}}/|d\mathbf{u}^{pq,\text{ideal,}\,\text{t}}|)$ & Eq.~(\ref{eq:duroll})        & 14.4 & 1.9 \\
\quad$\text{std}(\mathbf{t}^{p}ds^{p}/|d\mathbf{u}^{pq,\text{ideal,}\,\text{t}}|)$ & Eq.~(\ref{eq:tds})                   & 14.7 & 2.4 \\
\hline
Large strains ($-\varepsilon_{11}>0.01\%$) & & & \\
\quad$\text{std}(d\mathbf{u}^{pq,\text{centers, t}}/|d\mathbf{u}^{pq,\text{ideal,}\,\text{t}}|)$ & Eq.~(\ref{eq:durigid}) & 15.4 & 7.5 \\
\quad$\text{std}(d\mathbf{u}^{pq,\text{rigid, t}}/|d\mathbf{u}^{pq,\text{ideal,}\,\text{t}}|)$ & Eq.~(\ref{eq:centers})   &  9.7 & 7.9 \\
\quad$\text{std}(d\mathbf{u}^{pq,\text{roll}}/|d\mathbf{u}^{pq,\text{ideal,}\,\text{t}}|)$ & Eq.~(\ref{eq:duroll})        & 39.4 & 5.1 \\
\quad$\text{std}(\mathbf{t}^{p}ds^{p}/|d\mathbf{u}^{pq,\text{ideal,}\,\text{t}}|)$ & Eq.~(\ref{eq:tds})                   & 40.3 & 6.9 \\
\hline
\end{tabular}
\end{table}
The most evident finding is that all types
of contact motion are highly varied
and far more vigorous than that of the mean-field motion,
with quite large standard deviations
(to compare, the \emph{mean}
of the movements
$d\mathbf{u}^{pq,\text{centers, t}}/|d\mathbf{u}^{pq,\text{ideal,}\,\text{t}}|$
is approximately 1.0),
and the dispersion of the contact movements
increases with increasing strain.
Notably, the dispersion was considerably smaller among non-convex particles
than among spheres, and this reduced movement was most pronounced
for the rolling motions between non-convex particles.
Non-convex particles can touch at multiple points, and the contact
forces among multiple contacts produce resistive couples which
can suppress particle rotation.
The reduced dispersions of the sphere-cluster movements
is most marked in the dispersions of the
rolling movements $d\mathbf{u}^{pq,\text{roll}}$
and of the contact migrations $\mathbf{t}^{p}ds^{p}$,
which were much smaller
for the non-convex sphere-clusters than for the spheres
(the sphere values of 10 and larger are quite extreme).
\par
We also note that the dispersion in
the tangential contact movements,
$d\mathbf{u}^{pq,\text{rigid, t}}$, from which the contact forces
and bulk stress originate, was usually smaller than the dispersion
in the relative translations of the particle centers,
$\mathbf{u}^{pq,\text{centers, t}}$.
This result indicates that particle rotations
(i.e., the last term in Eq.~\ref{eq:durigid}) tend to offset
the relative translations of particles, reducing the relative
movements of the particles at their contacts and abating changes
in the tangential contact forces.
\section{Contact Longevity}
We now consider the longevity of contacts during bulk deformation,
a matter also addressed by \citeN{Hanley:2014a}.
Specifically, how long do grains typically remain in contact, and
at what rate are contacts created and expired during loading?
To answer these questions, we recorded a journal of all contact
creation and extinction events that had occurred
throughout the loading process.
Many contacts are brief collisional events
between rattler particles, not contributing to the load-bearing network
of the more persistent contacts.
We identified and removed these collisional contacts from the journal,
which were characterized by one of the two particles having only two
or fewer contacts at the time the contact was created and having
only one or zero contacts when the contact was parted.
This method of identifying contact duration
differs from that of \citeN{Hanley:2014a} who,
instead of creating a continuous journal, sampled the assembly
at strain increments of about $0.25\%$ 
(their $p^{'}$-constant simulations) and compared the contacts present
in these snapshots.
The current method is expected to yield smaller durations,
as the journal captures
multiple creation-extinction events that can occur across strains
of $0.25\%$ (or less) for the same particle pair.
With non-convex particles, we also distinguished between particle
\emph{neighbors} and particle \emph{contacts}, since
two neighboring (touching) particles can share multiple contacts.
\begin{table}
\centering\small
\caption{\small
         Longevities, creation rates, and separation rates of
         contacts during triaxial compression loading,
         in terms of elapsed strain $\Delta\varepsilon_{11}$.
         \label{table:longevities}}
\begin{tabular}{lccc}
\hline
 & \multicolumn{3}{c}{Particle shape}\\ \cline{2-4}
 &         & \multicolumn{2}{c}{Sphere-clusters}\\ \cline{3-4}
 & Spheres & Neighbors & Contacts\\
 \hline
Initial contacts & & &\\
\quad Mean longevity$^{\text{a}}$   & 0.019  & 0.021  & 0.012 \\
\quad Median longevity & 0.0068 & 0.0041 & 0.0031\\
\hline
Subsequent contact activity & & & \\
\quad Mean longevity   & 0.0016  & 0.021  & 0.0020 \\
\quad Median longevity & 0.00018 & 0.0024 & 0.00022\\
\quad Creation rate$^{\text{b}}$    & 680.    & 433.   & 545.   \\
\quad Separation rate  & 680.    & 433.   & 548    \\
\hline
\multicolumn{4}{l}{%
\parbox{3.80in}{\raggedright $\,^{\text{a}}$Longevities are in terms of
                elapsed strains $\Delta\varepsilon_{11}$.\\
                $\,^{\text{b}}$Rates are the numbers of
                contacts or neighbors created (or separated) divided by
                the total number of contacts or neighbors,
                and divided by the elapsed strain
                $\Delta\varepsilon_{11}$.}}\\
\hline
\end{tabular}
\end{table}
\par
The results are summarized in Table~\ref{table:longevities} for
the initial
set of contacts at the start of loading and also for all
contacts that
were subsequently created during triaxial compression.
These contact longevity values are in terms of the elapsed strain
$\Delta\varepsilon_{11}$
over which the particle pairs (or contacts) were touching.
The median longevity (half-life) of contacts within the sphere
assemblies was very brief.
Half of the contacts that were initially present at the start
of loading had separated after a strain of only 0.68\%;
whereas,
half of the contacts that had formed between spheres
during the subsequent loading had separated within an elapsed strain
$\Delta\varepsilon_{11}$ of only 0.018\%.
The median longevity of neighboring non-convex sphere-cluster particles is
somewhat more extended, as half of the sphere-clusters
that were established
during loading remained as neighbors for a strain of about 0.24\%.
Between these neighbors, however, the possibly multiple contacts
are more active, with a half-life strain of only 0.022\%.
\par
The ephemeral nature of contact is also seen in the
rapid rates of their
creation and separation (Table~\ref{table:longevities}).
The creation rate of $680$ for the sphere
assembly was computed by dividing the number
of contacts created during small
strain intervals by the average number of
contacts within these intervals and by the elapsed
strain $\Delta\varepsilon_{11}$.
This rate means that during an average interval of 1\% strain,
$6.8$ contacts were created for every current contact within the assembly,
but a corresponding $6.8$ contacts also separated during the same
1\% strain interval.
The activity among neighbors of sphere-cluster particles
is slightly more subdued:
for every neighboring pair at a given strain, $4.3$ neighbors formed
and $4.3$ neighbors separated during an average 1\% strain interval.
\par
These results are for slow quasi-static conditions rather than for
collision-driven rapid flow.
Yet even after the obviously collisional (rattler)
contacts had been eliminated,
these rates of contact creation and extinction demonstrate
the ephemeral, transient nature of contact interactions
during quasi-static loading.
\section{Persistence of Strong Contacts}
\begin{figure}
  \centering
  \includegraphics{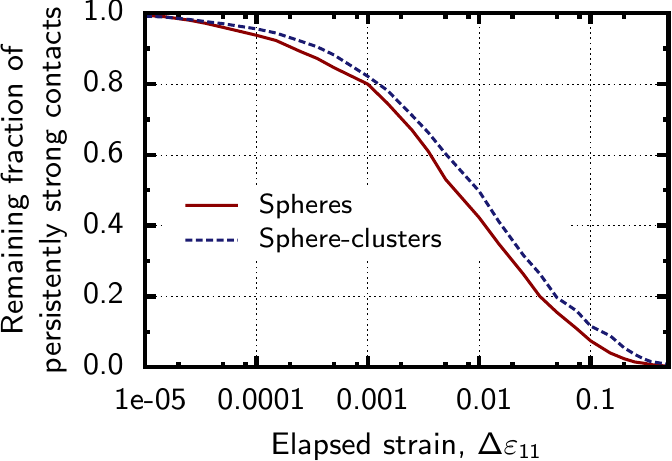}
  \caption{Diminution of an original subset of strong contacts,
           beginning
           at strain $-\varepsilon_{11}=0.01$, for the subsequent
           elapsed strain $\Delta\varepsilon_{11}$.
           \label{fig:Strong}}
\end{figure}

Recent attention has been given to the relative roles of the most
heavily and the most weakly loaded contacts during deviatoric loading.
\emph{Strong contacts}, those contacts with a normal force 
$f^{pq,\text{n}}$ that is
greater than the mean, predominantly support the deviatoric stress;
whereas, the stress attributed to weak contacts (with normal forces
smaller than the mean) is primarily isotropic \cite{Radjai:1996a}.
Close correlations have been found between the fabric of the strong
contact network and the evolving deviatoric stress \cite{Antony:2004b}.
We use results of DEM simulations to examine the \emph{persistence}
of this partition of the contacts between
the two sets --- strong and weak --- for assemblies of spheres and
of sphere-clusters.
With a pair of sphere-cluster particles,
which can have multiple contacts,
we used the total of the normal forces between the pair.
To measure persistence of the strong-contact partition,
we considered the set of all contacts at a
reference strain $-\varepsilon_{11}=1\%$ and then tracked the subset
of strong contacts that were
present at this original strain across the subsequent loading.
Those original strong contacts 
(at 1\% strain) that remained as strong contacts after an
elapsed strain $\Delta\varepsilon_{11}$ were considered ``persistently
strong'' across that strain interval.
Some contacts would transition between the strong and weak subsets,
and these contacts were not considered as being persistently strong.
Figure~\ref{fig:Strong}
shows the diminution of the persistently strong contacts across
a range of elapsed strains.
The results show that the partition of strong and weak subsets
was quite transitory: for example, fewer than half (42\%) of the original
strong contacts between spheres persistently remained
within the strong subset after the strain
had advanced by just 1\% ($\Delta\varepsilon_{11}=0.01$).
Only 7.5\% of the original strong subset
persisted across an elapsed strain
of 10\%.
The persistence of the strong-contact subset among
sphere-clusters, although brief,
is slightly more protracted than among
spheres.
\section{Movements at Distance and Force Chains}
\begin{figure}
  \centering
  \includegraphics{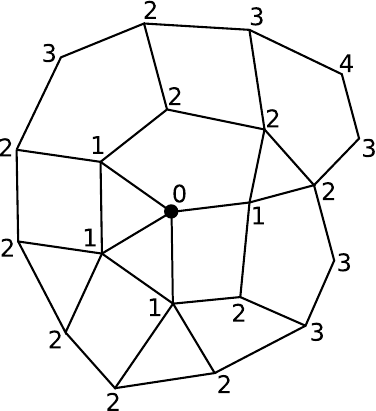}
  \caption{Discrete distances from single particle (node) ``$\bullet$''.
           \label{fig:distance}}
\end{figure}

As has been shown,
the relative movements of the centers of contacting particles,
$d\mathbf{u}^{pq,\text{centers}}$, are poorly
predicted by the mean-field movement
$d\mathbf{u}^{pq,\text{ideal}}$, as evidenced by the
large dispersions in Fig.~\ref{fig:scatter} and in Table~\ref{table:std1}.
The non-affine nature of the motions of neighboring grains
has been widely reported
\cite{Bagi:1993a,Calvetti:1997a,Kuhn:2003d,Tordesillas:2008a}.
If we instead consider pairs of particles that are
more remotely distanced from
each other, we would expect that the relative motions of these particles
would more closely conform with the mean-field deformation.
We might ask, at what distance (length scale) do particle movements
begin to approach compliance with the mean field?
To systematically investigate the
relative movements of particles that are
farther apart, we used a discrete measure
of the distance between pairs of particles that
were part of the assembly's load-bearing network.
A discrete, integer measure of distance has been introduced for
two-dimensional assemblies \cite{Kuhn:2003d},
which can be extended to three-dimensional
assemblies (see Fig.~\ref{fig:distance}).
Following principles of graph theory
\cite{Satake:1992a},
we constructed the adjacency matrix of the sphere assembly
at a reference strain, 
with one row and one column for each particle in
the load-bearing network
(i.e. excluding rattlers).
For this adjacency matrix,
1's were placed in locations $pq$ of contacting particles and
$\infty$'s were placed in all other matrix locations.
We then applied
the Floyd-Warshall algorithm to create a matrix $\mathbf{A}$ in
which an element $A_{PQ}$ is the shortest path
(discrete, integer distance)
between particles $P$ and $Q$.
For the $n$ particles in the load-bearing contact network,
this $n\times n$ symmetric matrix gives the distances between all
$n(n+1)/2$ combinations of particles $PQ$.
\subsection{Conformity of movements with mean deformation}
Our first use of the distance matrix $\mathbf{A}$ is determining
how consistently the relative movements of distant particles conform
with the mean deformation field.
In this regard,
we selected all particle pairs separated by a particular distance,
with discrete distances ranging from
$A_{PQ}=1$ (i.e., contacting neighbors) through 8.
For the 6400-particle assemblies, greater distances would
extend beyond the periodic boundaries.
As a measure of two particles' conformance with the mean deformation,
we used a normalized inner product:
\begin{equation}
  \chi^{PQ} = d\mathbf{u}^{PQ,\text{centers}}\cdot
              d\mathbf{u}^{PQ,\text{ideal}}
              / |d\mathbf{u}^{PQ,\text{ideal}}|^{2}
  \label{eq:chiPQ}
\end{equation}
where $d\mathbf{u}^{PQ,\text{centers}}$ and 
$d\mathbf{u}^{PQ,\text{ideal}}$ are computed with
Eqs.~(\ref{eq:centers}) and~(\ref{eq:pureComponents})
for the distant particles $P$ and $Q$.
Particle movements in perfect conformity with the mean-field
deformation have $\chi=1$; movements in counter-conformity
have $\chi=-1$; and movements that are orthogonal to the
mean-field movement have $\chi=0$.
\par
\begin{figure}
  \centering
  \includegraphics{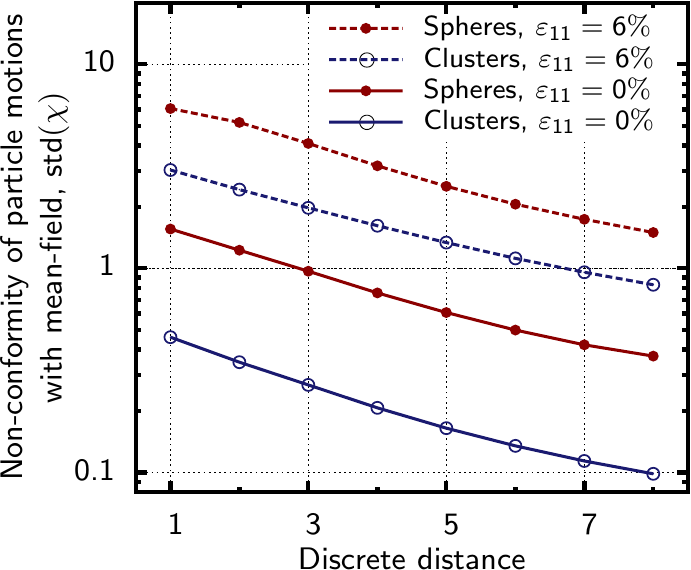}
  \caption{Inconsistency of contact motions with the
           mean-field deformation during triaxial
           compression of the sphere assembly, expressed with the 
           standard deviation of fluctuations from the mean
           field.  The discrete distance
           is the number of contacts separating two
           particles. \label{eq:conformance}}
\end{figure}

Figure~\ref{eq:conformance} gives the
dispersions (fluctuations) in $\chi$ at two strains,
reported as standard deviations in $\chi$:
at the start of loading and at strain
$-\varepsilon_{11}=6\%$ for assemblies of
spheres and of sphere-clusters.
To place these results in perspective,
the \emph{mean} values of $\chi$ were very close
to 1.0 at both strains, at all distances,
and for both particle shapes:
\emph{on average}, the motions of particle pairs were consistent with
the mean deformation.
The average inconsistency (dispersion or fluctuation) in conformance
is indicated by the standard deviation
of $\chi$, measured at different distances.
The figure shows that conformance improves with the distance
(discrete separation) between particles.
Moreover, particle movements conform more closely to the mean-field
deformation at the start of loading $\varepsilon_{11}=0$
than at the larger strain.
Even at the start of loading, however, conformance is far from
perfect, with the standard deviation $\text{std}(\chi)$ equal
to 1.56 for contacting sphere pairs and improving to 0.37 for
sphere pairs that are
separated by 8 contacts (at the ends of chains of 9 particles).
Again, these standard deviations can be compared with the
mean of 1.0,
demonstrating a relatively large degree of deviation of
the particle movements from the affine field for
particles separated by as many as 8 contacts.
\par
We note a much better conformance with the mean-field deformation
for the
non-convex sphere-clusters than for the spheres.
At the start of loading, the standard deviation of $\chi$ was 0.46
for the contacting sphere-clusters: although large, this dispersion
is much smaller than the 1.56 of contacting spheres.
Indeed,
sphere-clusters at distances of 5 or greater conform
fairly closely with the mean-field
deformation at the start of loading
($\text{std}(\chi)$ values less than 0.16).
The conformance is very poor, however, at larger strains,
and even the lnog-range motions do not approach the affine ideal.
\subsection{Persistence of force chains}
We used similar methods to evaluate the longevity of
chains of heavily loaded particles (so-called "force chains"),
which are an ubiquitous feature of the load-bearing
fabric of granular materials
\cite{Drescher:1972a,Thornton:1986a,Majmudar:2005a}.
Simulations have shown that force chains are
non-stationary, evolving structures that alter the local
topologic arrangements of particles and
the load-bearing fabric
\cite{Walker:2012a,Oda:1998b}.
We identified these force chains by using a weighted graph,
in which each
edge (contact) was assigned its inverse normal force,
$1/f^{pq,\text{n}}$, as a weight
(for non-convex particles, the denominator is the sum
of normal forces at the multiple contacts between a particle pair).
A weighted adjacency matrix
was then populated with these inverse
contact forces for the contacting
particles but with $\infty$'s placed in other matrix locations. 
The Floyd-Warshall algorithm was used to compute
a "nearest weighted distance" matrix $\mathbf{F}$, in which each element
$F_{PQ}$ is the sum of the inverse forces of all contacts
along the contact chain
between two particles
$P$ and $Q$.
We defined and identified the dominant force chains with two criteria.
Recognizing the significant length of force chains,
we first considered the subset of pairs $PQ$ that
were separated by 6 contacts (that is, chains of 7 particles),
found by
scanning the discrete distance matrix $\mathbf{A}$ for the
condition $A_{PQ}=6$.
The representative intensities of forces within these $PQ$
chains were characterized by the harmonic means
$\overline{f}$ of their normal forces:
\begin{equation}
  \overline{f}^{PQ,\text{n}} =
  A_{PQ}/F_{PQ} =
  A_{PQ} \left( \sum_{pq\in PQ} 1/f^{pq,\text{n}} \right)^{-1}
\end{equation}
where $pq$ are the contacts in the chain between particles
$P$ and $Q$.
Note that different metrics were used in computing distance
matrices $\mathbf{A}$ and $\mathbf{F}$,
and different particle chains (geodesics) can
be associated with the corresponding elements $A_{PQ}$ and $F_{PQ}$,
a subtlety
that was ignored in our calculation of mean force
$\overline{f}^{PQ,\text{n}}$.
As the second criterion,
we identified those particle pairs within the first subset
(pairs at distance~6) 
whose force intensity $\overline{f}^{PQ,\text{n}}$
ranked among the largest 10\%
of all pairs within this subset.
The two criteria provided a systematic means of identifying
the dominant force chains at a given instant in the triaxial loading process.
(Other methods have also been used for identifying force chains, such as
\citeNP{Zhang:2001A})
\par
\begin{figure}
  \centering
  \includegraphics{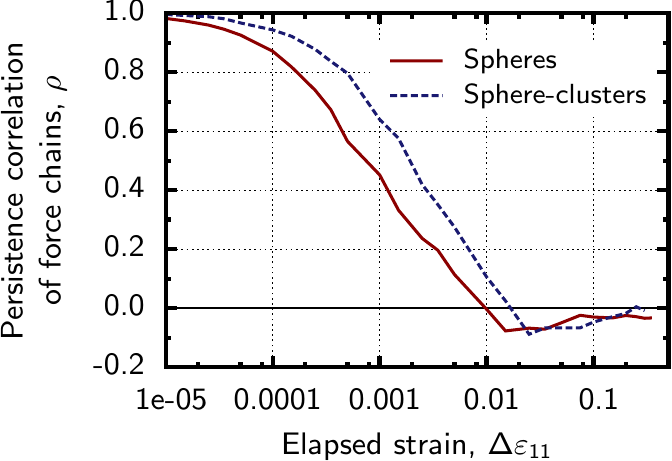}
  \caption{Persistence of force chains that were present
           at strain $-\varepsilon_{11}=6\%$ across
           subsequent strains $6\% + \Delta\varepsilon_{11}$
           \label{eq:ChainsPersist}}
\end{figure}

To investigate the temporal persistence 
of these dominant force chains,
we compared the set of force chains at the
reference strain $-\varepsilon_{11}=6\%$
with the set of force chains
at several later strains of $6\%+\Delta\varepsilon_{11}$.
That is, we compared the set of
force chains present near the peak stress with
the set of force chains that had developed
after an elapsed strain $\Delta\varepsilon_{11}$.
We considered the union of these two sets of $PQ$ pairs and found the
correlation of their force intensities
$\overline{f}^{PQ,\text{n}}$ at the two strains.
Because the mean stress
could differ at the two strains
(and, with it, the average contact force), we compared
the \emph{rankings} of the mean forces $\overline{f}^{PQ,\text{n}}$
among force chains at the two strains,
using Spearman's rank correlation coefficient $\rho$
as a measure of the persistence of the dominant force chains.
If the rankings of particle pairs were the same at the two strains,
then $\rho=1$; but if they had opposite rankings, then $\rho=-1$.
Figure~\ref{eq:ChainsPersist} gives this measure of the persistence
of force chains that were originally present
at the reference strain $-\varepsilon_{11}=6\%$ for the subsequent
strains. 
The increment $\Delta\varepsilon_{11}$ ranged from an almost
instantaneous difference of 0.001\% to the considerable interval of 40\%,
well beyond the peak stress and near the critical state.
\par
For spheres and for sphere-clusters,
the correlation $\rho$ is nearly reduced to zero for elapsed strains
of 1\% and larger,
indicating that force chains present at strain 6\% had largely been
replaced by other force chains at strain 7\%.
At a smaller elapsed strain of 0.1\%, the correlation $\rho$ of
0.6 (or less) demonstrates that force chains are fairly ephemeral
structures that are constantly being formed and transformed
across small time scales.
\section{Conclusions}
\par
Progress in engineering mechanics follows a sequence
of five endeavors:
observation, understanding, explanation, modeling, and prediction.
Much of the research on granular materials (including the current work)
has been in the first three areas,
and progress on the first has advanced the furthest:
examining and cataloging granular behaviors at various scales.
A largely unresolved problem in granular mechanics is the accurate
modeling and prediction of bulk
mechanical behaviors from micro-scale analyses.
These mechanical behaviors include the stress (strength) that is
attained after a given deformation
history and the incremental stiffness for a given particle packing or state.
In this regard,
the greatest attention has been given to predicting
incremental stiffness at small strains,
since the movements of contacting or nearby particles
during the start of loading
conforms most closely to an affine assumption.
The simplest models are based upon two-particle systems or upon systems
that include the first and second shells
of particles around a central particle.
The current observations do not provide much encouragement
for such modeling approaches:
deviations from the affine condition are severe,
even if the third or more distance shells of particles are considered.
At large strains, the effort appears to be futile, since
contact movements become complex,
darting sequences of sliding and rolling motions that are far
from the affine field,
and particle arrangements are quite fleeting.
\par
Some consoling results do appear among the data, however.
Although individual force chains are transient features that
are weakened and replaced during strain intervals of as small as 1\%,
their presence is fairly resilient for increments of up to 0.01\%.
As such, the micro-scale modeling of force chains could
become the basis of predicting incremental stiffness and strength.
The recent work of Tordesillas and her coworkers has greatly advanced
three of the five activities:
understanding, explaining, and modeling the origins of force chains
(for example, \citeNP{Tordesillas:2009a}).
The partitioning of contacts into weak and strong contacts
also exhibits some of the same qualities as force chains,
since this partition is fairly resilient for strain increments
of as large as 0.1\%. 
If future research can improve understanding
and explain the origin of strong contacts
(e.g., which features exist in the neighborhoods
of the most heavily loaded contacts and particles?),
models based upon the strong-weak
partition are promising candidates for predicting stiffness and strength.
\par
The current study also shows that non-convex particles
behave in a more regular fashion than spheres.
The movements of non-convex sphere-clusters are less varied,
their strong-weak partition is more temporally resilient,
their movements are not as far removed from an affine field,
and their force chains are more persistent.
These simulation results apply to rather smooth sphere-clusters,
which have a shape that is only modestly non-convex and less
irregular than that
of most natural grains.
In this regard, micro-scale models
of simpler sphere particles using near-affine kinematics might
give reasonable predictions for natural materials,
even though the model might yield poor predictions for sphere assemblies!
As another possible approach,
analyses of force chains and strong contacts could be extended to include the
effects of non-convexity and non-sphericity in the particle shapes.
\par
A more unusual approach for estimating granular strength and stiffness
is to eschew the notions of regularity
and order and to
assume that the material is thoroughly disordered, even maximally
so.
This approach has been taken by the author in modeling fabric and strength
at the critical state \cite{Kuhn:2014a}.
Although the results are encouraging, they do not yet address
the incremental stiffness
or the evolution of deviatoric stress in the pre-peak regime.
Answers in these areas may require a synthesis of the modeling approaches
mentioned above:
the use of non-affine kinematic fields and non-convex particle
shapes with attention to
the dominant meso-scale
structures (e.g., strong particles or force chains), while incorporating
maximum disorder principles.
\section*{\normalsize Acknowledgement}
The author great-fully acknowledges productive discussions
of this work with Dr.~WaiChing Sun.
%
%
%

\end{document}